\documentclass[aps, prd, a4paper, twocolumn, nofootinbib, showpacs, amsfonts]{revtex4}

\usepackage{amssymb,latexsym}
\usepackage{amsmath, amsthm}
\usepackage{amscd}
\usepackage{times}
\usepackage{epsfig}
\usepackage{psfrag}
\usepackage{graphicx}

\newcommand{\bea}{\begin{eqnarray}}
\newcommand{\eea}{\end{eqnarray}}
\newcommand{\be}{\begin{equation}}
\newcommand{\ee}{\end{equation}}

\begin{document}

\title{Electromagnetic counterparts to gravitational waves from black hole mergers and\\
naked singularities}

\author{Daniele Malafarina}
\email{daniele.malafarina@nu.edu.kz}

\affiliation{Department of Physics, Nazarbayev University, 010000 Astana, Kazakhstan}

\author{Pankaj S. Joshi}
\email{psj@tifr.res.in}

\affiliation{Tata Institute of Fundamental Research, Homi Bhabha Road, Colaba, Mumbai 400005, India}

\date{\today}

\begin{abstract}
We consider the question here whether the proposed electromagnetic counterpart of the gravitational wave signals in binary black hole coalescence may be due to the appearance of a `short lived' naked singularity during the merger. We point out that the change in topology that the spacetime undergoes during the merger can cause the appearance of a naked singularity. In case some matter, in the form of a small accretion disk, is present in the surroundings of the black hole system then the emitted luminosity during the merger would allow to distinguish the scenario where the naked singularity forms from the scenario where the horizon exists at all times.
In fact the emitted luminosity spectrum would be much higher in the case where a naked singularity forms as opposed to the `pure' black hole case.
We suggest that the presence of such a transient naked singularity will explain the high luminosity of an electromagnetic counterpart during the merger much more easily.
\end{abstract}


\maketitle

\section{Introduction}
The recent claim by the LIGO collaboration of the detection of gravitational waves from a black hole binary merger (GW150914)
\cite{LIGO}
was soon accompanied by the claim by the FERMI collaboration of the detection of a possible electromagnetic counterpart
\cite{Connaughton}.

However, if two black hole horizons were to merge in vacuum
then no such emission of electromagnetic waves should be allowed.
One possible explanation of the electromagnetic counterpart (if confirmed) is given by the existence of an accretion
disk surrounding the binary system
\cite{Loeb}.
The proposed mechanism here is very similar to the standard collapsar model.
It was suggested by
\cite{Lyutikov}
how to possibly explain the short duration of the observed gamma ray burst if little amount of matter should be present in the disk.

In the present paper our purpose is to suggest another possible explanation for the existence of electromagnetic counterparts to gravitational waves detected from binary black hole mergers. These may occur due to the change of topology of the spacetime that accompanies the black hole merger, and which results in the formation of a `short lived' naked singularity.
In fact, if the black hole merger and the topology change that accompanies such an event gives rise to a naked singularity,
may be for limited but short time, then the accretion disk around the same could in principle emit electromagnetic radiation much more efficiently than the black hole case, and would explain such high radiative powers. Also, the strong negative pressures that could appear close to the naked singularity, which is a
quantum object that is classically described by the spacetime singularity, which are due to quantum effects, may allow for the ejection of part of the mass of the original objects.

We note that any accretion disk surrounding a binary black hole
merger, such as the ones considered in
\cite{Loeb} and \cite{Meszaros},
would have an accretion power determined by the presence of the horizon and the existence of an innermost stable circular orbit (ISCO).
On the other hand, the appearance of a transient naked singularity
and any small accretion disk around the same would give away a much higher luminosity more efficiently than the corresponding situation in which a black hole horizon exists at all times. The measured luminosity spectrum could
then be used as a possible observational test to constrain such hypothesis.

It was Wheeler who suggested that singularities in solutions of Einstein's equations may indicate a window on quantum-gravitational effects
\cite{Wheeler}.
It has been known for a long time that topology changes in a spacetime manifold are linked to causality violations and/or singularities
\cite{Geroch},
and that singularities generically occur as a result of such topology changes
\cite{Tipler}.
Furthermore it was shown in
\cite{Joshi}
that topology change in a spacetime implies the occurrence of a naked singularity.
Therefore, if a singularity that is not covered by a horizon should
form during a process that leads to change in the topology of the spacetime, then
some quantum effects occurring in its vicinity could in principle have a detectable
observational signature and implication.
A binary black hole merger is a typical situation in which such topology change would occur.
Then if a naked singularity forms for a limited time during the merger, it is possible that quantum effects arising in the strong field region when the system is not covered behind a horizon, may have observational
consequences.
Also, from an observational perspective, the comparison between the properties of accretion disks around black
holes and naked singularities was studied in
\cite{JMN}.
There it was seen that the emission spectra are significantly different for these objects, the naked singularity accretion disk being much more effective radiator, and also therefore they are in principle observationally distinguishable from each other.


The paper is organized as follows. In section \ref{topology} we review the basic features of changes in topology of a spacetime in connection with naked singularities
and we discuss the implications of such changes for the coalescence of two black holes. In section \ref{accretion} we outline how the emission of radiation, when an accretion disk is present near a naked singularity in the system can help explain the proposed electromagnetic counterpart to GW150914. Finally comments and remarks are presented in section \ref{discussion}.

\section{Topology change, naked singularities and black hole mergers} \label{topology}
It was Geroch who showed that any spacetime containing two spacelike hypersurfaces with different topologies must have either closed timelike curves or singularities
\cite{Geroch}.
The argument was refined by Tipler who showed that by imposing the weak energy condition it is possible to ensure that there cannot be any topology change without the presence of a spacetime singularity
\cite{Tipler}.
The visibility or otherwise of such singularities was then investigated in connection with the Cosmic Censorship Hypothesis, and one of us showed that naked singularities in terms of past incomplete nonspacelike geodesics do arise as a consequence of topology change
\cite{Joshi}.

There are several approaches that one can adopt in order to interpret such singularities. For example, topology change can be viewed as a purely classical phenomenon provided one allows to extend general relativity to include degenerate metrics, and therefore the singularities associated with them can be considered to be in some sense `mild'
\cite{Horowitz}.
On the other hand, a very powerful and fruitful idea is to interpret the singularities that arise in classical GR as an indication of breakdown of the theory and therefore use them as milestones towards a viable theory of quantum gravity.
A fully quantum-gravitational approach should take care of such singularities, removing the undesired infinities without breaking the causal structure of the manifold. Nevertheless, even without a full theory of quantum-gravity, one can use a semi-classical approach to quantum corrections to show how these would affect the strong field regime by using purely classical tools, {\it i.e.} by solving the usual Einstein field equations (see for example
\cite{semiclassical}).

Then, if naked singularities can be viewed as a window on quantum-gravity, the appearance of one in the merger of two black holes may possibly produce a powerful explosive phenomenon (such as a GRB) whose nature would be intrinsically quantum-gravitational. Such an event would be a cosmic laboratory where one could observe and test quantum-gravity.

The coalescence of two black holes is a typical situation in which a topology change occurs.
A binary black hole system is characterized by an initial topology of two disjoint event horizon hypersurfaces, and the final topology (after the merger) is a single null hypersurface which is the resultant event horizon of the merged black hole. In the simplest case of two Schwarzschild black holes merging, the initial topology is that of two disjoint circles evolving, and final topology is given by $S^2 \times\mathbb{R}$.
Then if a naked singularity were to form during the merger event,
a window would open on the ultra-high density compact object located at the point of merger. For example, overspinning of a Kerr solution could provide a mechanism for the formation of a naked singularity.
Note that this process would seem to violate the black hole area theorem. This is consistent with fact that the black hole area theorem assumes that the spacetime is weakly asymptotically simple and empty, which is cosmic censorship assumption, that is violated at the time of the merger due to the topology change. In other words, for the black hole area theorem to hold one must assume that cosmic censorship is valid. On the other hand a violation of cosmic censorship, as in the case illustrated here, allows for the area theorem not to hold, at least temporarily.

In this connection, a specific possibility that may occur is, when two Kerr black holes of spin very close to unity collide and merge, then at least momentarily a Kerr geometry with $a>1$ could develop (where $a$ is the adimensional spin parameter given by $a=J/M$). This would then eventually settle to a Kerr black hole by radiating away a sufficient amount of angular momentum through a suitable mechanism. Such a scenario would be entirely consistent with the topology change theorems mentioned above.

\begin{widetext}

\begin{figure}[h]
\centering
\begin{minipage}{.45\textwidth}
\centering
\includegraphics[scale=0.25]{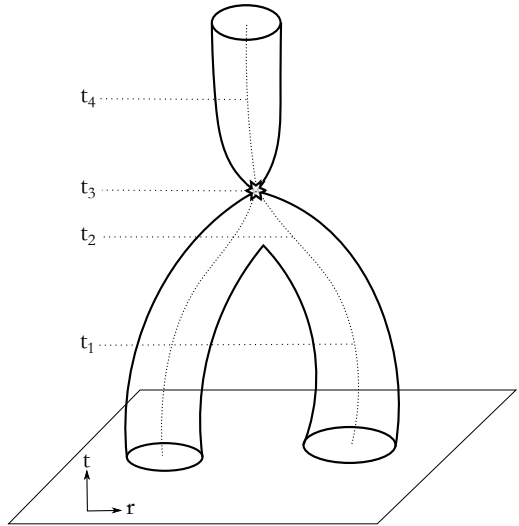}
\end{minipage}
\hfill
\begin{minipage}{.45\textwidth}
\centering
\includegraphics[scale=0.25]{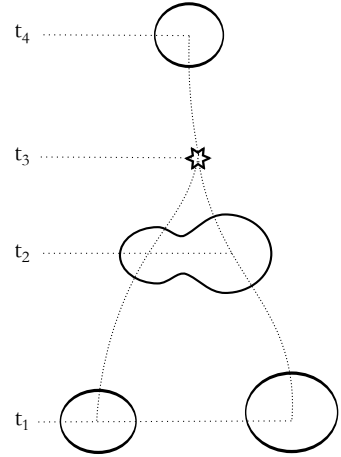}
\end{minipage}
\caption{Qualitative depiction of the merger of two black holes with formation of a transient naked singularity. On both panels, at the time $t=t_1$ the black holes are separated. Thick lines represent the horizons and dotted lines the central singularities. At the time $t=t_2$ the two horizons merged into a single horizon, the singularities are still separated. At the time $t=t_3$ the singularities merge forming a temporary overspinning Kerr spacetime. There is no horizon. After $t=t_3$ the excess angular momentum is radiated away and the system settles to a Kerr black hole, as at the time $t=t_4$.}
\label{fig1}
\end{figure}

\end{widetext}

Whether it is possible to create a Kerr solution with spin parameter greater than one is one of the key unanswered questions of black hole physics today. Several astrophysical black hole candidates are known with spin parameter close to unity. Is it possible to add enough angular momentum to such objects in order to destroy the event horizon? In
\cite{Jacobson}
it was suggested that infalling test particles carrying angular momentum can indeed overspin a nearly extremal Kerr black hole and turn it into a naked singularity.
On the other hand in
\cite{Barausse}
it was shown that the inclusion of the gravitational self-force could prevent the above scenario from happening.
All these examples involve test particles carrying angular momentum and are restricted by simplifying assumptions on the initial set up.
A completely different scenario is that of the merger of two black holes of comparable size, both carrying angular momentum and with an arbitrary initial configuration.
The merger of two such black holes would have to carry away most of the excess angular momentum for the final configuration to settle to a Kerr solution with $a<1$.
Simulations have shown that the merger of two black holes can carry away enough angular momentum to respect the Kerr bound
\cite{Rezzolla}.
However, there are several parameters that come into play when describing a merger in full and one cannot exclude at present that an intermediate nakedly singular phase can form from some initial configurations (see Figure \ref{fig1}), as suggested by the topology change considerations.

We should also recall here that the ringdown phase in the detected GW150914 event should not be necessarily taken as a strong indication of the formation of the event horizon necessarily, as the same signal may be obtained by a small ringed compact object
(see \cite{Cardoso}).
Therefore it is worth asking the question whether a binary merger such as that associated with GW150914 could produce a transient quantum-gravitational phenomenon, in which the central object, classically described by a naked singularity, is not covered by the horizon for a short finite time. Allowing for this possibility, we are now interested in investigating what kind of electromagnetic signal could come from vicinity of such an object.


\section{Accretion onto a naked singularity} \label{accretion}
In the observed GW150914 the merger carried away an estimate of three solar masses in gravitational waves. In the case where the singularities of the colliding black holes are hidden behind the horizon at all times this amount of energy cannot come from the irreducible mass of the black holes that is possibly concentrated in the singularity. The energy radiated away in gravitational waves must come from the binding energy of the system, loss of angular momentum and/or from the matter in the accretion disk. Accretion disks around black hole candidates are known for being not very massive, typically of the order of $0.1M_{\odot}$, and therefore the amount of energy that they could provide is not enough. Further to this, for a binary system like the one that originated GW150914 the surrounding medium is expected to be almost entirely devoid of matter. On the other hand, as we shall see below, a naked singularity is a much more efficient engine to convert the gravitational energy of the infalling particles into radiation. The repulsive forces that may occur near the singularity as a result of quantum effects could also possibly strip some mass away from the central object to eject it in the surrounding medium, thus providing a source of matter for the accretion disk.

In
\cite{JMN}
we studied the properties of accretion disks around naked singularities. For steady state accretion, the luminosity for an accretion disk around a compact source can be written as
\be
\mathcal{L}_{\rm acc}=\epsilon \dot{m} c^2 ,
\ee
where $\dot{m}$ is the accretion rate of the mass in the disk and
$\epsilon=1-E_{\rm ISCO}$ is the efficiency,
which is related to the binding energy per unit mass of circular orbits evaluated at the ISCO ($E_{\rm ISCO}$)
(see for example \cite{Thorne}).
For a Schwarzschild black hole the ISCO is located at $6M$, $a$ is equal to zero and $\epsilon=0.057$. In the limiting case of a Kerr black hole with $a$ going to one the ISCO shrinks to $r_{\rm ISCO}=M$ and the efficiency reaches the maximum allowed value for a black hole of $\epsilon=0.42$. As we can see, the efficiency depends on the location of the ISCO. For a naked singularity the stable circular orbits could in principle extend all the way to the center and the efficiency can be as high as one hundred percent. If we interpret the singularity as a measure of our ignorance on quantum gravity, we can assume that it must be replaced by a finite sized exotic compact object of Planck size. Then the ISCO would not be exactly zero and thus the emitted power would be lower than a `pure' naked singularity but still much higher than that of a black hole. Similarly, the time interval during which the source is not covered by the horizon depends upon the specific nature of the central object. It may reduce to an instant in the limit of a naked singularity in the collapse frame but would appear longer and finite to faraway observer.

From these considerations we see that knowing the mass of the final configuration and assuming a scenario such as the one presented here we could use the measured luminosity to constraint the size and `lifespan' of the exotic central object.

Given a theoretical model for a compact source one can evaluate $L_{\rm acc}$ in order to compare the observed spectrum with the predicted one.
For circular accretion disk the net luminosity for the observer at infinity is given by the integral over $\ln r$ of
\be
\frac{d\mathcal{L}}{d \ln r}=4\pi r\sqrt{-g}E\mathcal{F} ,
\ee
where $r$ is the radial coordinate in the equatorial plane, $\sqrt{-g}$ is the determinant of the three dimensional metric outside the compact source (black hole or not) in the equatorial plane of the disk, $E$ is the energy of the test particle of the disk at a distance $r$ from the center and $\mathcal{F}$ is the radiative flux emitted by the disk at a distance $r$
(see \cite{JMN} and references therein).
The flux is in turn given by
\be\label{F}
\mathcal{F}=-\frac{\dot{m}}{4\pi \sqrt{-g}}\frac{\omega_{,r}}{(E-\omega L)^2},
\int_{r_{\rm ISCO}}^r (E-\omega L)L_{,r}dr
\ee
where $\omega$ and $L$ are the angular velocity and angular momentum of the test particle in the disk.
For a general stationary and axially symmetric metric of the form
\be
ds^2=g_{tt}dt^2+2g_{t\phi}dtd\phi+g_{rr}dr^2+g_{\theta\theta}d\theta^2+g_{\phi\phi}d\phi^2,
\ee
test particles in the accretion disk are confined to the equatorial plane $\theta\simeq \pi/2$ and their energy, angular momentum and angular velocity can be written as
\bea
E&=& -\frac{g_{tt}-g_{t\phi}\omega}{-g_{tt}-2g_{t\phi}\omega-g_{\phi\phi}\omega^2},\\
L&=& \frac{g_{t\phi}-g_{\phi\phi}\omega}{-g_{tt}-2g_{t\phi}\omega-g_{\phi\phi}\omega^2},\\
\omega&=&\frac{-g_{t\phi}+\sqrt{(g_{t\phi,r})^2-g_{tt,r}g_{\phi\phi,r}}}{g_{\phi\phi,r}}.
\eea
Then the location of the ISCO can easily be evaluated from the second derivative of the effective potential for particles in the disk
\be
V_{\rm eff}=\frac{E^2g_{\phi\phi}+2ELg_{t\phi}+L^2g_{tt}}{g_{t\phi}^2-g_{tt}g_{\phi\phi}}-1.
\ee

Note how the above quantities depend on the specific nature of the spacetime metric coefficients. Therefore the ISCO is determined by the metric. By looking at equation \eqref{F} it is not difficult to see that for $r_{\rm ISCO}$ going to zero $\mathcal{F}$ diverges, therefore suggesting that in a realistic scenario the singularity must be replaced by a finite Planck sized exotic compact object.


Every object in space has a maximum luminosity $\mathcal{L}_{\rm Edd}$, called the Eddington luminosity, beyond which radiation pressure overcomes gravity.  Once the accretion flow reaches $\mathcal{L}_{\rm Edd}$ the material outside the object stops falling inwards and gets pushed away from it. It is possible for an accreting compact object to have a luminosity that exceeds its Eddington limit depending on the specific features of the accreting fluid. Nevertheless it is generally believed that the Eddington limit may not be surpassed by much.
In
\cite{Lyutikov}
it was pointed out how the proposed electromagnetic counterpart to GW150914 detected by FERMI, that lasted approximately one second, had a luminosity that exceeds by ten orders of magnitude than that expected for a $60M_{\odot}$ black hole. Therefore, while keeping the assumption that the two detections were caused by the same event, to make sense of this high luminosity one can either invoke an extremely high accretion rate on a black hole or one can postulate a lower and more realistic accretion rate on an exotic compact object.

Note that a transient naked singularity forming from the merger would require an accretion rate of the order of $10^{-5}M_{\odot}/sec$ to produce a GRB of luminosity
\be
\mathcal{L} \simeq 10^{49}\frac{erg}{sec} ,
\ee
which is therefore much smaller than that of black hole. This would allow or require for a much less massive accretion disk to be present in the surrounding medium at the time of the event to accomplish the electromagnetic signal. A naked singularity event of the duration of one second would allow the compact source to emit the electromagnetic burst before settling down to a final black hole configuration. The duration of the GRB would then be related to the characteristic lifetime of the naked singularity. This in turn would be given by the time required by the horizon to collapse plus the time necessary to shed away the excess angular momentum. The time scale of the latter would then be related to the natural scale of the underlying quantum-gravity theory.
Then the measurement of the duration of the burst could provide information on the nature of the exotic compact object at the center of the process.


\section{Discussion}\label{discussion}
The detection of the gravitational wave signal GW150914 from the LIGO observatory has opened a new era in astrophysics. The possibility of the existence of an electromagnetic counterpart to the signal as detected by the FERMI observatory has been suggested. If the existence of electromagnetic counterparts to gravitational waves during binary black hole merger was to be confirmed that would raise a lot of questions about the nature of the sources of gravitational waves. This would be
especially so if the measured luminosity of the event would prove to be much higher than the Eddington luminosity for the binary black hole system.

In the present paper we asked the question whether a transient naked singularity appearing during the merger could be used as a source for the observed electromagnetic signal. Given our ignorance on the quantum nature of the exotic compact object that is classically described by the singularity we cannot construct a full model describing the process. For example we do not know if the matter in the accretion disk is coming from the medium outside the original black holes or if the repulsive pressures originating close to the singularity can eject part of the mass of the original object and add it to the accretion disk.
On the other hand we have seen how the measurement of the luminosity of these events could provide observational constraints on the nature of the compact object that forms during the merger. Therefore, while more observations are needed in order to better understand the nature of the phenomenon, at present it is worth asking whether it is possible to use such observations to confirm or rule out the possibility of the creation of a transient naked singularity during the coalescence.

If hypothesis such as the one above were to be proven correct the study of these events would then provide a measurable cosmic laboratory where to test and observe quantum-gravity at work.

{\em Acknowledgement:} The authors thank Prashant Kocherlakota, Jun Qi Guo, Andrzej Krolak, Abraham Loeb, K. I. Nakao, and Ramesh Narayan, for useful discussions on these themes. DM would like to thank TIFR for support and hospitality while this work was done. This article was supported by the Nazarbayev University Social Policy grant.

\end{document}